\documentclass[12pt]{article}

\usepackage[letterpaper, portrait, margin=1in]{geometry}
\usepackage[utf8]{inputenc}
\usepackage{float}
\usepackage{authblk}
\usepackage{enumitem}
\usepackage{multirow}
\usepackage{caption}
\usepackage[style=ieee]{biblatex}
\usepackage{mathtools}
\usepackage{amssymb}
\usepackage{physics}
\usepackage[group-separator={,}]{siunitx}
\allowdisplaybreaks

\newcommand{\Universe}{\mathbb{R}^3}
\newcommand{\Vector}[1]{\vb{#1}}
\newcommand{\Point}[1]{\vb{\lowercase{#1}}}
\newcommand{\Manifold}[1]{\uppercase{#1}}
\newcommand{\Boundary}[1]{\partial \Manifold{#1}}

\DeclareMathOperator{\overhangs}{overhangs}

\newcommand{\Overhangs}[2]{\Point{#1} \,\overhangs\, \Point{#2}}

\newcommand{\VerticalUnit}[1]{\vu{k}_{\Point{#1}}}
\newcommand{\PositionVector}[2]{\Vector{r}_{\Point{#1} \leftarrow \Point{#2}}}
\newcommand{\PositionUnit}[2]{\vu{r}_{\Point{#1} \leftarrow \Point{#2}}}
\newcommand{\Height}[2]{z_{\,\Point{#1} \leftarrow \Point{#2}}}
\newcommand{\AngleOfElevation}[2]{\theta_{\Point{#1} \leftarrow \Point{#2}}}
\newcommand{\AngleReducedHeight}[2]{z\,'_{\!\Point{#1} \leftarrow \Point{#2}}}

\DeclareMathOperator{\dom}{dom}
\DeclareMathOperator{\sub}{sub}
\DeclareMathOperator{\jut}{jut}
\DeclareMathOperator{\rut}{rut}

\newcommand{\Dominance}[1]{\dom(\Point{#1})}
\newcommand{\Submission}[1]{\sub(\Point{#1})}
\newcommand{\Jut}[1]{\jut(\Point{#1})}
\newcommand{\Rut}[1]{\rut(\Point{#1})}

\newcommand{\latitude}{\phi}
\newcommand{\longitude}{\lambda}
\newcommand{\slope}{\alpha}
\newcommand{\aspect}{\beta}

\newcommand{\Rotation}[2]{R_{#1}(#2)}

\title{Beyond Elevation:\@ New Metrics to Quantify the Relief\\ of Mountains and Surfaces of Any Terrestrial Body}
\author[]{Kai Xu}
\affil[]{Yale University}
\date{}

\begin{document}

\graphicspath{{./figures/}}

\maketitle

\begin{abstract}

Elevation has long been the standard for quantifying the relief of mountains and other landforms on Earth and beyond. Nevertheless, elevation has its limitations. By itself, a location's elevation reveals little about its vertical position relative to its surroundings, especially for seabed and extraterrestrial features. Furthermore, on planets and asteroids without a sea level, the zero-elevation datum is defined rather arbitrarily, making elevation values rather meaningless without also considering relative elevation differences. In light of these factors, this paper introduces new topographic measures based purely on gravity and the actual planetary surface, rather than on an arbitrary datum. Unlike elevation, the so-called datumless measures---with the names of \textit{dominance}, \textit{jut}, \textit{submission}, and \textit{rut}---each describe a different aspect about a location's vertical position relative to its surroundings. They can be used to objectively compare the relief of mountains and other landforms, including across different planets. The datumless measures may be of interest to planetary scientists seeking universally standardized measures of relief, Earth scientists seeking to correlate geomorphological properties with natural phenomena, as well as mountaineers and hobbyists seeking new ways to describe the wondrous landscapes of this universe.

\end{abstract}

\section{Introduction}

Central to the field of topography is elevation, which is widely used to quantify relief on Earth and other planets.\footnote{For conciseness, a \textit{planet} will refer to any astronomical object with a terrestrial surface, including moons and asteroids.} Yet, despite its ubiquity, elevation has its limitations. On Earth, the vertical datum, from which elevation is based, has been subject to numerous changes based on arbitrary and inconsistent conventions, resulting in minor elevation differences depending on the datum used \parencite{datum-1}. The greater problem, however, arises on planets without a sea level, where the datum is defined especially arbitrarily due to the lack of an obvious terrestrial feature to base it on, hence the lack of a universally standardized way to quantify relief \parencite{datum-2}. Furthermore, when quantifying relief on such planets, the elevation of a point is rather meaningless on its own, deriving most of its value when compared to the elevation of other points. Therefore, elevation is better described as a reference frame for comparing height differences, rather than as an absolute measure of relief in and of itself. That is not to disregard elevation altogether, as it is invaluable for many tasks such as mapping, geolocation, and climatology. However, when it comes to the distinct task of quantifying topographic relief, it is beneficial to think outside the datum.

In light of these factors, this paper introduces a universally consistent framework for quantifying relief that does not rely on a datum. The so-called datumless measures are based purely on physically meaningful phenomena without reliance on arbitrary parameters.

Following this introduction, the preliminaries required for understanding the datumless measures are introduced. Preliminaries include a formal definition of the planetary surface, as well as a framework for describing the spatial relationship between two points on the planetary surface.

Next up are the datumless measures themselves, which describe various aspects about the vertical position of a location relative to local terrain (unlike elevation, which describes vertical position relative to an imaginary datum). The \textit{dominance} measure describes how high a location rises above its large-scale surroundings, and is perhaps most similar to elevation of all the measures introduced. The \textit{jut} measure describes how sharply or impressively a location rises above its immediate surroundings (usually from the bottom of a major mountain face or a neighboring valley), factoring in both height above surroundings and steepness. Inspired by the omnidirectional relief and steepness (ORS) measure \parencite{spire-measure}, which was designed to quantify the visual impressiveness of mountains and other protruding landforms, jut achieves the same objective while being easier to visualize and understand. Meanwhile, the \textit{submission} measure describes how high the large-scale surroundings of a location rise above the location itself. A mountaintop with a submission of 0 is called a \textit{dominant point}, and can be thought of as a non-arbitrarily defined ``local high point.'' Finally, the \textit{rut} measure describes how sharply or impressively the surroundings of a location rise above the location itself, considering both the height and angle of elevation of the surroundings above the location. It can be used to identify cities and other non-mountainous locations with visually impressive mountain backdrops.

The final technical section of this paper explains how the datumless measures can be computed. This section also showcases the values of the datumless measures at various places on Earth, Moon, Mars, and Vesta.

\section{Preliminaries}

\subsection{Datumless Definition of the Planetary Surface}

This subsection introduces a definition of the planetary surface that does not rely on a reference ellipsoid, and is based instead on meaningful physical concepts. The provided definition is conducive to both digital representation and mathematical abstraction.

\subsubsection{Planet as a Manifold-With-Boundary}

Let the universe be represented by 3-dimensional space, or \(\Universe\). Every location relative to the planet of interest can be mapped to a point\footnote{The term \textit{point} will refer to a position vector, which provides more notational flexibility than an ordered triple.} in \(\Universe\). This paper uses an Earth-Fixed, Earth-Centered (ECEF) coordinate system. However, as long as the relative positions of points on the planet are represented properly, the absolute coordinate system does not matter.

Let the planet of interest be represented by \(\Manifold{m}\), a theoretically infinite set of points in \(\Universe\), i.e., \(\Manifold{m} \!\subset \Universe\). A point in \(\Universe\) is in \(\Manifold{m}\) if and only if its location is occupied by physical matter that the planet of interest is composed of. Depending on the intentions of measurement, \(\Manifold{m}\) can be defined to include or exclude matter corresponding to features such as buildings, vegetation, water bodies, and permanent ice and snow.

\begin{figure}[H]
\centering
\frame{\includegraphics[width=0.5\columnwidth]{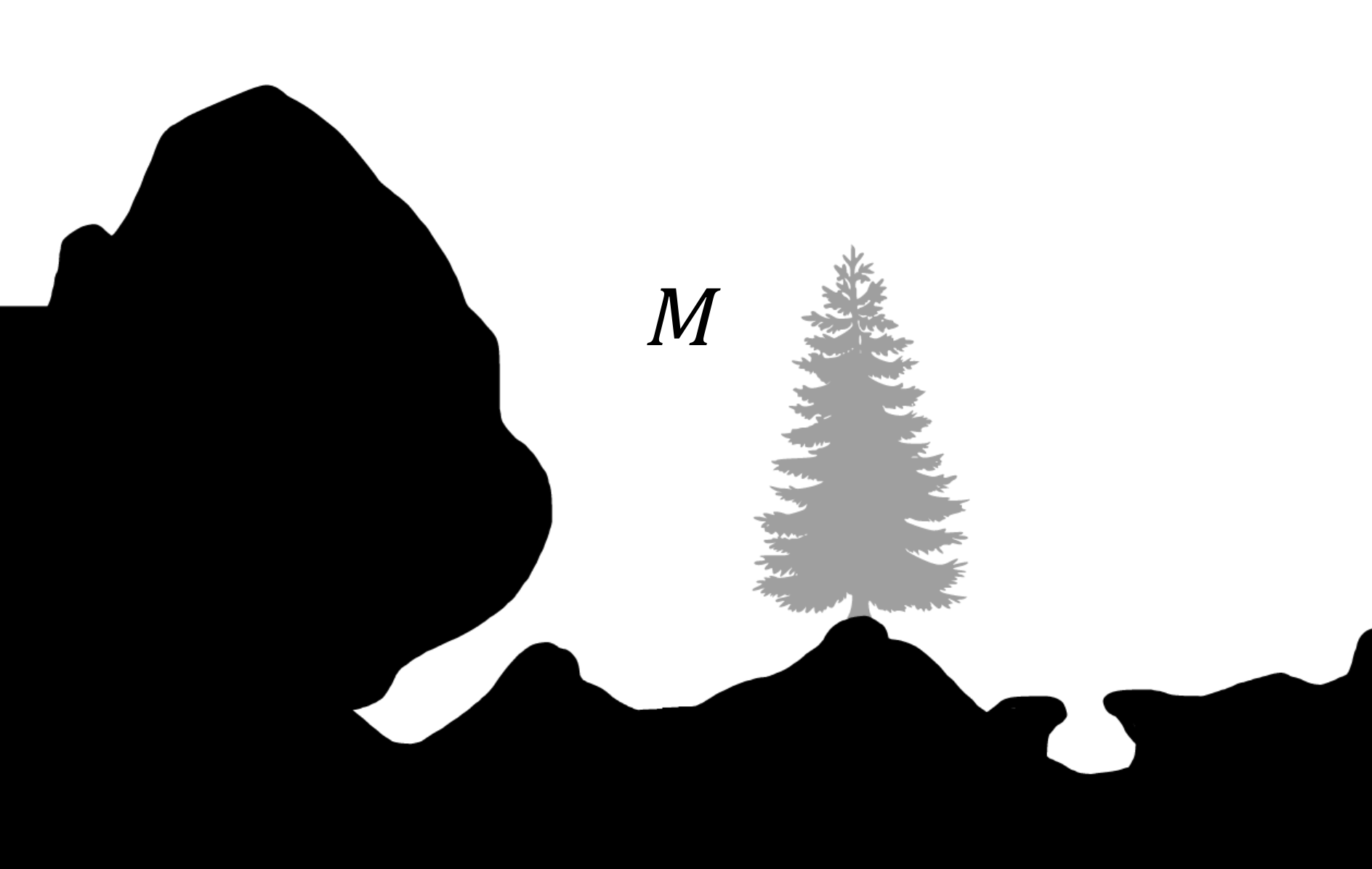}}
\caption{
Points in \(\Manifold{m}\) are shaded in black. Vegetation is excluded in this particular definition of \(\Manifold{m}\).
}
\end{figure}

Effectively, \(\Manifold{m}\) is a manifold-with-boundary---more specifically, a 3-manifold with a 2-dimensional boundary. The boundary of \(\Manifold{m}\), denoted by \(\Boundary{m}\), comprises of all points in \(\Manifold{m}\) that are infinitesimally close to a point in \(\Universe\) that is not in \(\Manifold{m}\). The boundary can be thought of as the parts of a planet that are directly exposed to the atmosphere (or outer space, if an atmosphere does not exist).
\[
\Boundary{m} = \qty{\Point{p} \in \Manifold{m} \mid \qty(\exists\, \Point{q} \in \qty{\Universe \setminus \Manifold{m}} : \abs{\,\Point{p} - \Point{q}\,} = \varepsilon)}
\]
where \(\varepsilon\) denotes an infinitesimally small quantity.

\(\Boundary{m}\) has a fractal-like texture all the way down to the microscopic level. It is not useful for topographic applications, as it contains points on cave walls and underneath overhangs, which are not captured by most surface models such as digital elevation models (DEMs). One need not worry about representing \(\Manifold{m}\) or \(\Boundary{m}\) digitally---they are merely conceptual intermediaries used to define the planetary surface.

\begin{figure}[H]
\centering
\frame{\includegraphics[width=0.45\columnwidth]{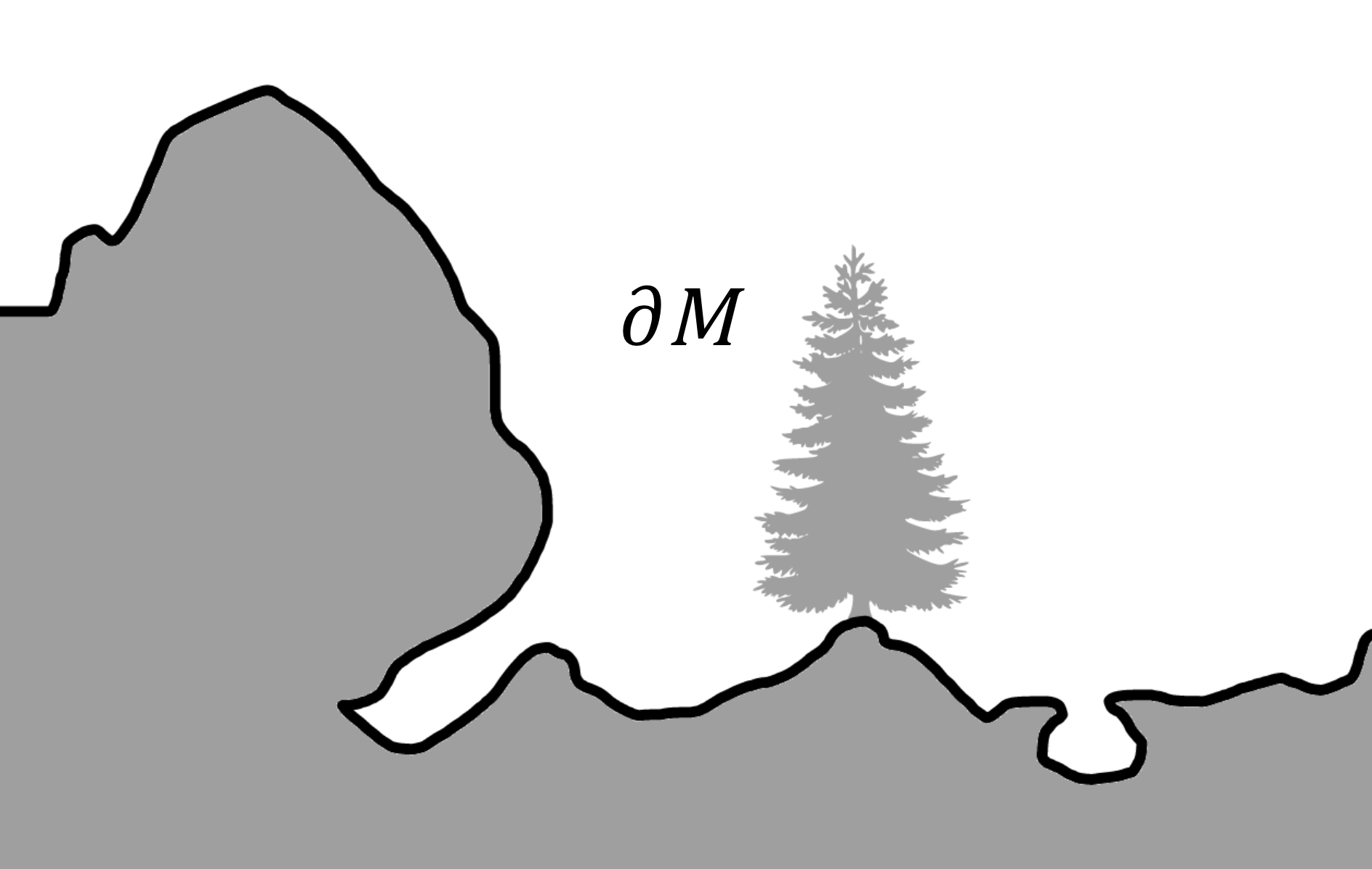}}
\caption{
Points in \(\Boundary{m}\) are outlined in black.
}
\end{figure}

\subsubsection{Planetary Surface}

A planet's gravity field can be depicted as surfaces of constant gravity potential surrounding the planet---also known as \textit{equipotential surfaces}---to which the direction of gravity is always perpendicular to. \textit{Plumblines} are defined as curves running perpendicular to the equipotential surfaces, that are therefore tangent to the direction of gravity along their lengths. Roughly speaking, an object will fall along its plumbline.

To introduce terminology used in defining the planetary surface, point \(\Point{q}\) \textit{overhangs} point \(\Point{p}\) if and only if \(\Point{q}\) has a higher gravity potential than \(\Point{p}\) and also lies on the same plumbline as \(\Point{p}\).

\begin{figure}[H]
\centering
\includegraphics[width=0.5\columnwidth]{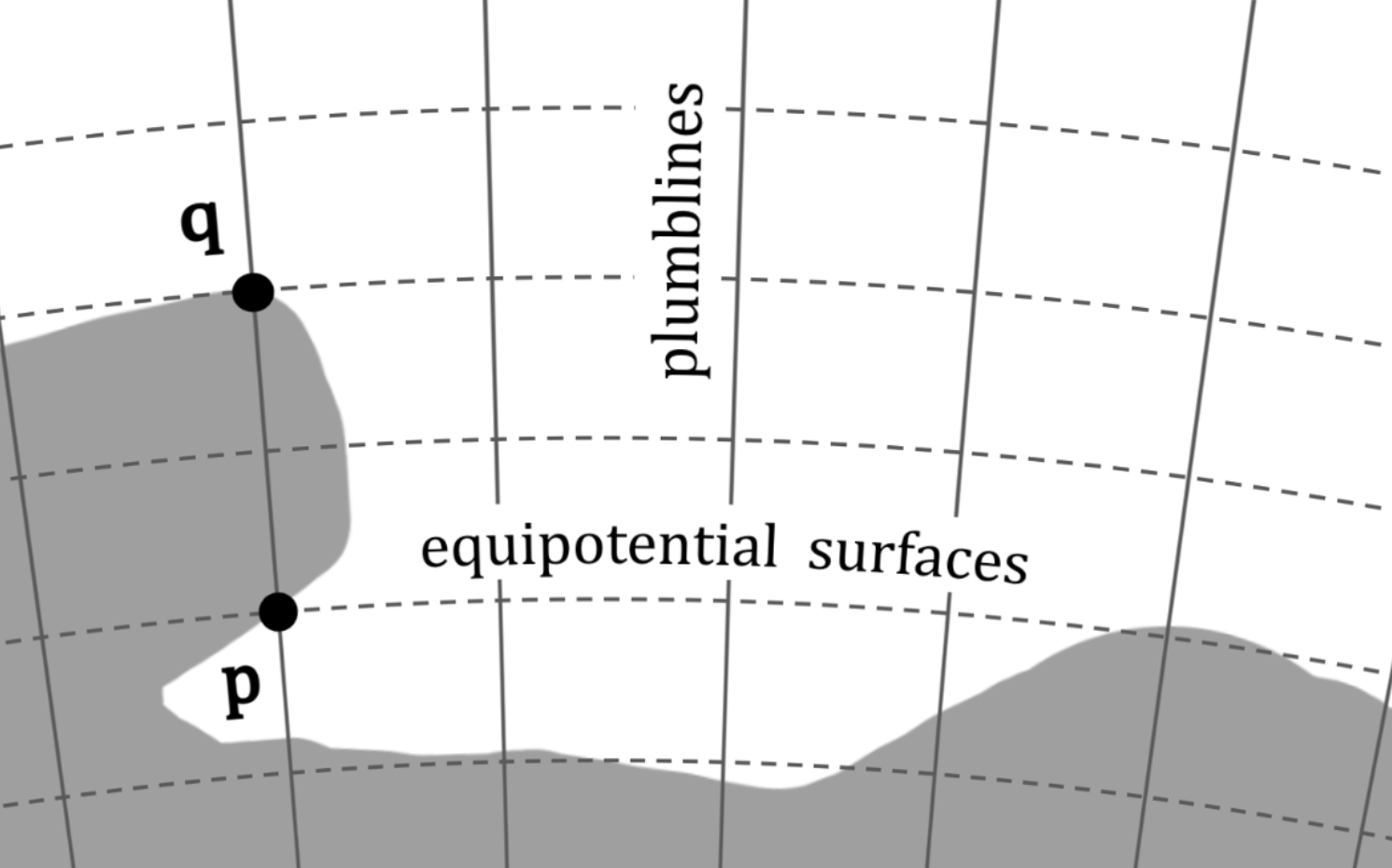}
\caption{
Point \(\Point{q}\) overhangs point \(\Point{p}\) in this diagram.
}
\end{figure}

The \textit{planetary surface}, denoted by \(\Manifold{s}\), is the subset of \(\Boundary{m}\) consisting of all points in \(\Boundary{m}\) that are not overhung by another point in \(\Manifold{m}\). The planetary surface can be thought of as all parts of a planet that are directly exposed to falling raindrops.
\[
\Manifold{s} = \qty{\Point{p} \in \Boundary{m} \mid \qty(\nexists \Point{q} \in \Manifold{m} \mid \Overhangs{q}{p})}
\]

\(\Manifold{s}\) is an infinite set of points, meant to represent a perfect topographic model of infinite resolution. In practice, \(\Manifold{s}\) can simply be represented by a DEM.\footnote{In this case, the usage of elevation in a DEM is non-arbitrary, as it is for determining the objective locations of points on the planetary surface, rather than as a subjective measure of relief.} On Earth, depending on how \(\Manifold{m}\) is defined and which surface model is used, the planetary surface can either be \textit{wet}, including water surfaces in its set of points, or \textit{dry}, including underwater bathymetry instead. In this paper, \(\Manifold{s}\) and all derived measurements should be assumed as wet unless stated otherwise.

\begin{figure}[H]
\centering
\frame{\includegraphics[width=0.45\columnwidth]{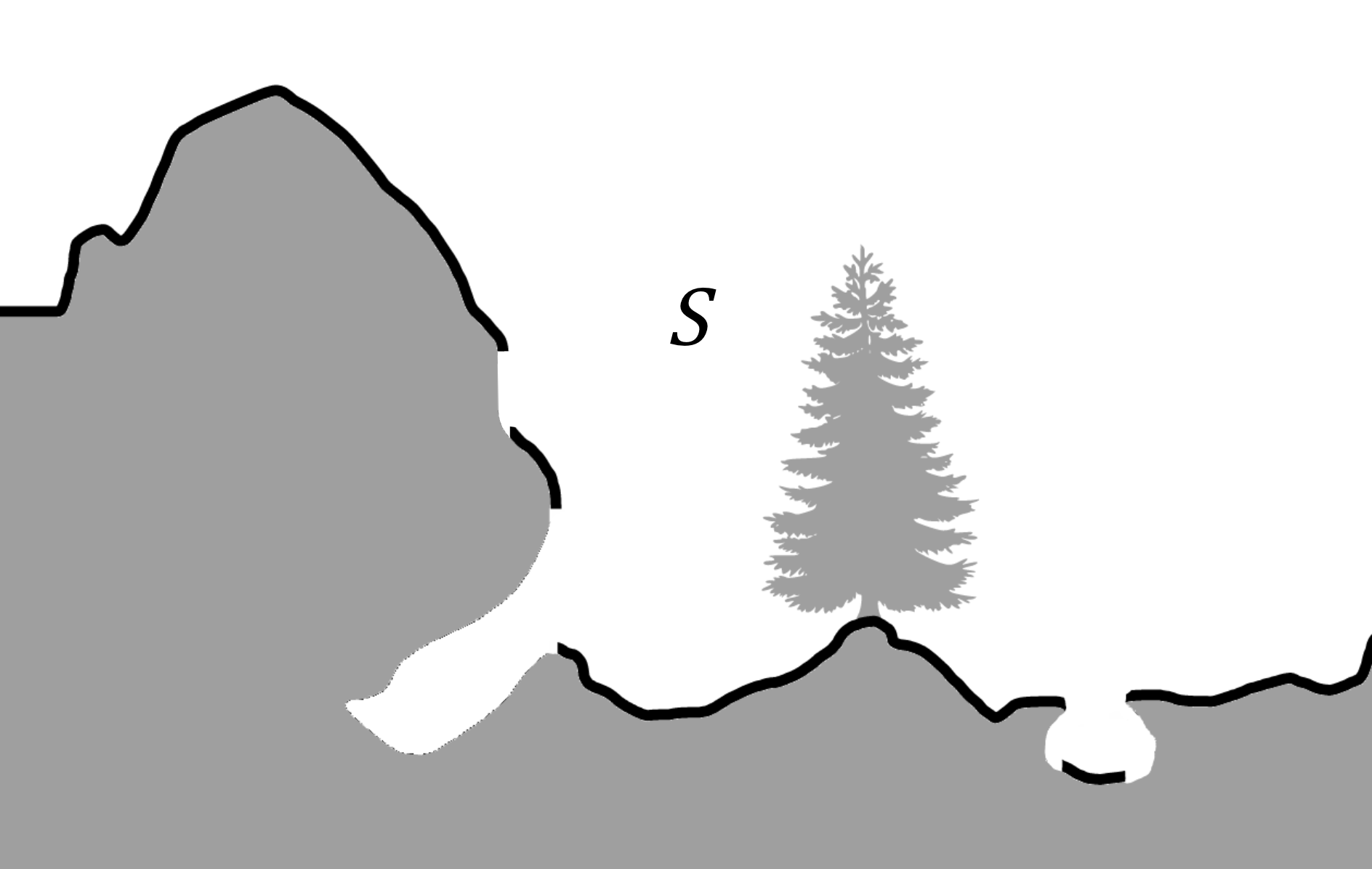}}
\caption{
Points in \(\Manifold{s}\) are outlined in black.
}
\end{figure}

\subsection{A Point in the Reference Frame of Another Point}

The concepts in this subsection are best understood through the analogy of an observer looking at point \(\Point{p}\) from point \(\Point{q}\), ignoring viewshed obstructions. For instance, \(\Point{p}\) could be the summit of a mountain, and \(\Point{q}\) could be the location of an observer at the foot of the mountain.

\subsubsection{Vertical Unit Vector and Horizontal Plane}

Let the \textit{vertical unit vector} of \(\Point{q}\), denoted by \(\VerticalUnit{q}\), be the vector of magnitude 1 that points \textit{opposite} the direction of gravity at \(\Point{q}\).\footnote{It is preferred to define the vertical direction as aligning with the local direction of gravity (rather than aligning with the planet's center of mass) to ensure perceptually accurate measurements on irregularly shaped asteroids.} The vertical unit vector represents the upwards direction of an observer at \(\Point{q}\).

Let the \textit{horizontal plane} of \(\Point{q}\) be the flat plane passing through \(\Point{q}\) that \(\VerticalUnit{q}\) is perpendicular to. Another point \(\Point{p}\) is \textit{above} the horizontal plane of \(\Point{q}\) if it is on the same side of the plane as \(\VerticalUnit{q}\) points towards, \textit{below} the horizontal plane if it is on the opposite side, and \textit{on} the horizontal plane if it touches it. The horizontal plane can be thought of as the ``flat eye level'' of an observer standing at \(\Point{q}\), except this plane intersects the planetary surface where the observer is standing, rather than literally being at eye level.

\subsubsection{Height Above the Horizontal Plane}

Let the \textit{position vector} of \(\Point{p}\) with respect to \(\Point{q}\), denoted by \(\PositionVector{p}{q}\), be the vector pointing from \(\Point{q}\) to \(\Point{p}\), i.e., the difference between their coordinates:
\[
\PositionVector{p}{q} = \Point{p} - \Point{q}
\]
The arrow subscript notation as used above denotes an attribute of \(\Point{p}\) in the reference frame of \(\Point{q}\). This notation will continue to be used.

The \textit{height} of \(\Point{p}\) above the horizontal plane of \(\Point{q}\), denoted by \(\Height{p}{q}\), is equal to the scalar projection of \(\PositionVector{p}{q}\) onto the vertical unit vector of \(\Point{q}\).
\[
\Height{p}{q} = \PositionVector{p}{q} \vdot \VerticalUnit{q}
\]

Point \(\Point{p}\) is \textit{above} the horizontal plane of \(\Point{q}\) if \(\Height{p}{q}\) is positive, \textit{on} the horizontal plane if \(\Height{p}{q}\) is 0, and \textit{below} the horizontal plane if \(\Height{p}{q}\) is negative. Due to planetary curvature, the height of one point above the horizontal plane of another is not equal to the elevation difference between the two points. For instance, even though Mount Everest has a much higher elevation than the Dead Sea, Mount Everest is over 1000 kilometers below the horizontal plane of the Dead Sea.

\subsubsection{Angle of Elevation}

The \textit{angle of elevation} of \(\Point{p}\) above the horizontal plane of \(\Point{q}\), denoted by \(\AngleOfElevation{p}{q}\), is the signed angle between \(\PositionVector{p}{q}\) and the horizontal plane of \(\Point{q}\). It is positive if \(\Point{p}\) is above the horizontal plane of \(\Point{q}\), negative if \(\Point{p}\) is below the horizontal plane, and 0 if \(\Point{p}\) is on the horizontal plane.
\[
\AngleOfElevation{p}{q} = \arcsin(\PositionUnit{p}{q} \vdot \VerticalUnit{q})
\]
where \(\PositionUnit{p}{q}\) is the position unit vector, defined as follows:
\[
\PositionUnit{p}{q} = \frac{\PositionVector{p}{q}}{\abs{\PositionVector{p}{q}}}
\]
The value of \(\AngleOfElevation{p}{q}\) is undefined if \(\Point{p} = \Point{q}\).

\begin{figure}[H]
\centering
\frame{\includegraphics[width=0.6\columnwidth]{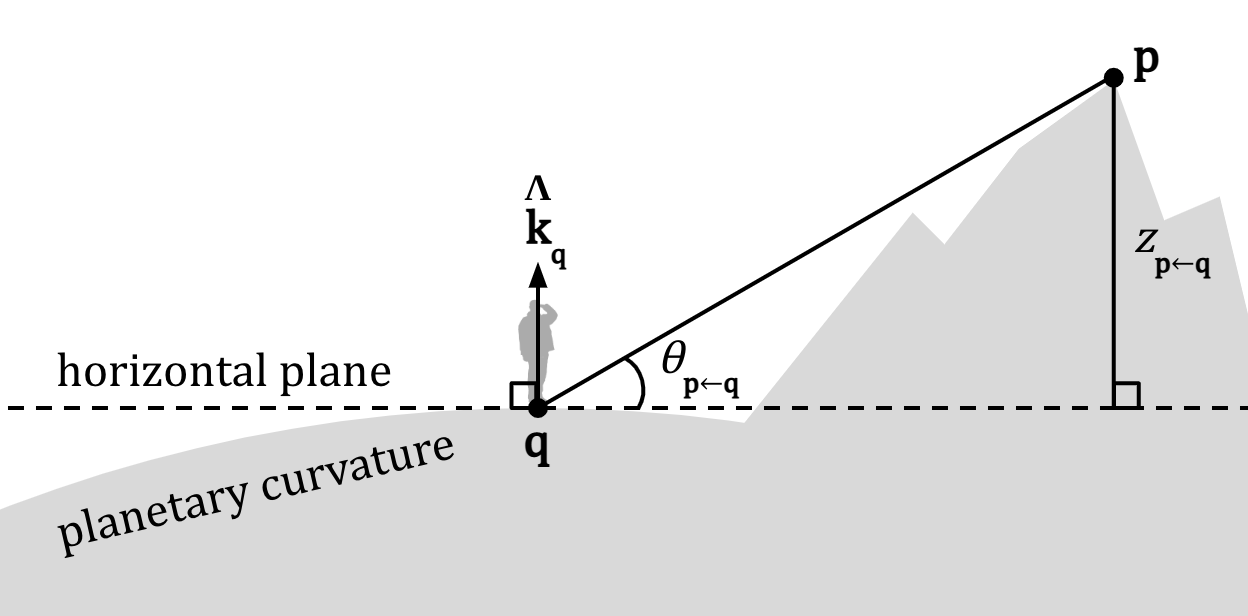}}
\caption{
Point \(\Point{p}\) in the reference frame of point \(\Point{q}\).
}
\end{figure}

\subsubsection{Angle-Reduced Height}

Let the \textit{angle-reduced height} of \(\Point{p}\) above \(\Point{q}\), denoted by \(\AngleReducedHeight{p}{q}\), be equal to the following:
\[
\AngleReducedHeight{p}{q} = \Height{p}{q} \, \abs{\sin \AngleOfElevation{p}{q}}
\]
and if \(\Point{p} = \Point{q}\)\,, let \(\AngleReducedHeight{p}{q} = 0\).

The magnitude of angle-reduced height describes how sharply point \(\Point{p}\) gains its relief with respect to \(\Point{q}\). Its magnitude increases with both a greater height difference and a steeper angle of elevation. Meanwhile, the sign of angle-reduced height describes which side of the horizontal plane of \(\Point{q}\) that \(\Point{p}\) is on.

\begin{figure}[H]
\centering
\frame{\includegraphics[width=0.4\columnwidth]{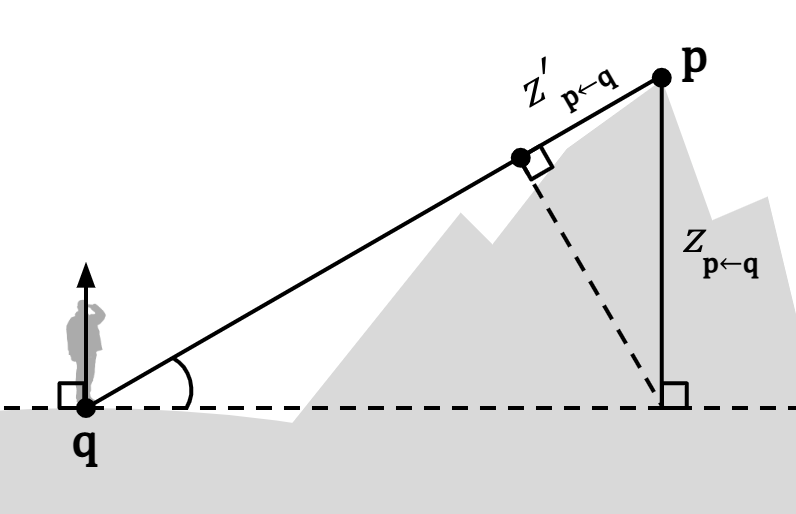}}
\caption{
Geometric diagram of angle-reduced height.
}
\end{figure}

Subjectively, angle-reduced height provides a simple formula that correlates with how visually impressive point \(\Point{p}\) appears to an observer at \(\Point{q}\). For instance, the impressiveness of a mountain is a result of not only how high it rises, but also the angle at which it does, hence why mountains tend to appear more impressive from close-up. To capture this phenomenon, angle-reduced height is equal to height above the horizontal plane when \(\AngleOfElevation{p}{q}\) is \ang{90} or \ang{-90} (akin to a vertical cliff). However, the less steep the angle of elevation, the lower the value of \(\abs{\sin(\AngleOfElevation{p}{q})}\) is and the more the expression for \(\AngleReducedHeight{p}{q}\) gets reduced.

Earl and Metzler \parencite{spire-measure} was the first to present a formula that specifically captures the visual impressiveness of a point as observed from another, serving as an inspiration for this paper. In their paper, the formula \(H \frac{\theta}{\ang{90}}\) was used, where \(H\) is the absolute value of the elevation difference between two points and \(\theta\) is the angle of elevation on a flat Earth model, where the elevation difference and geodetic distance between two points are treated as the legs of a right triangle. Angle-reduced height presents an improvement upon this formula, as it is easier to visualize geometrically, defined for angles greater than \ang{90}, and does not require a datum by means of elevation.

\begin{figure}[H]
\centering
\includegraphics[width=0.7\columnwidth]{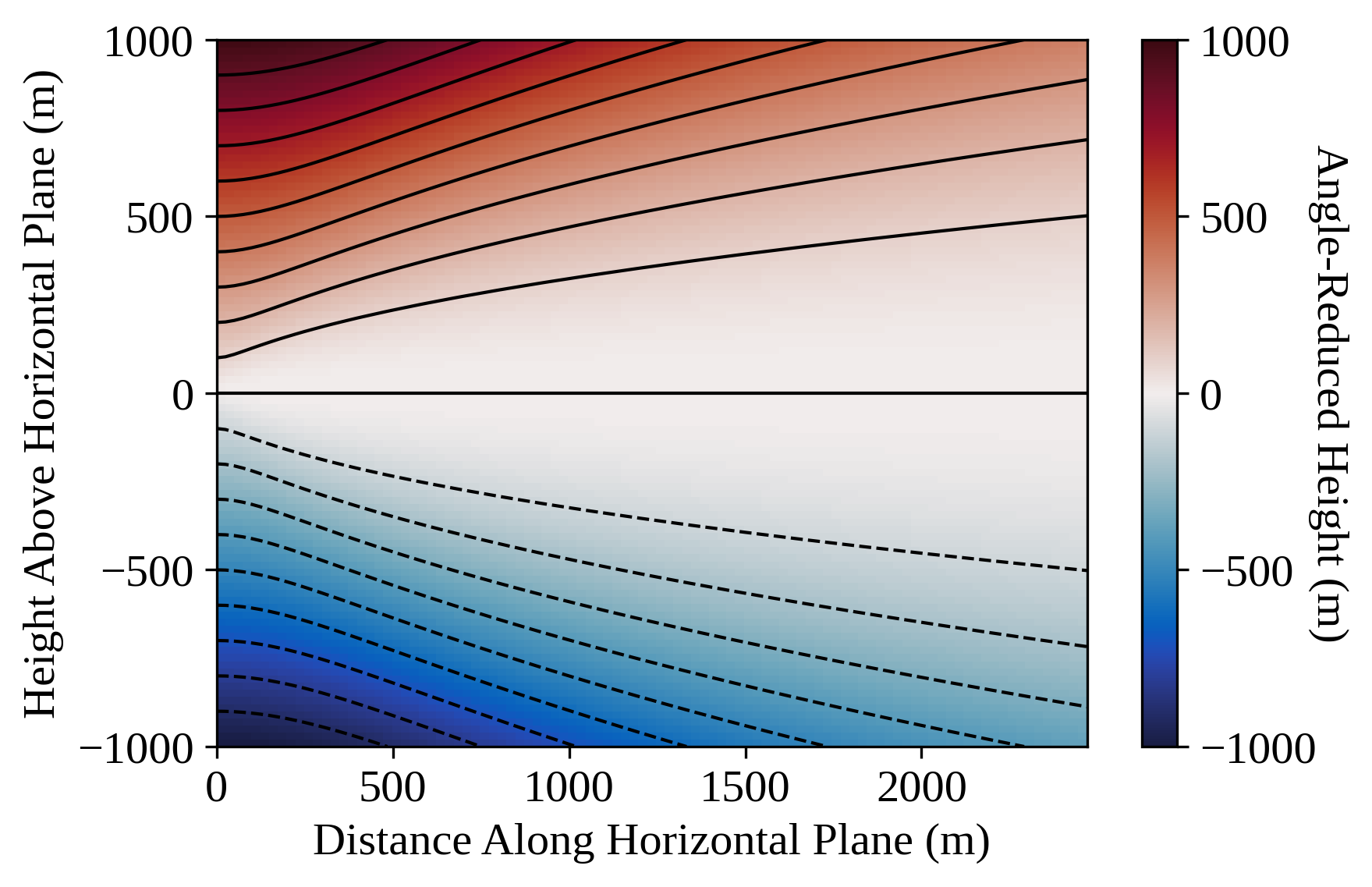}
\caption{
Contour plot of angle-reduced height. If the origin of this graph represents the location of an observer and the horizontal axis represents their horizontal plane, points on the same contour line (and those of the same color) are considered equally impressive to the observer. This graph is created in Matplotlib \parencite{matplotlib}. The red-white-blue colormap used in this and subsequent visualizations is adapted from cmocean colormaps \parencite{cmocean}.
}
\end{figure}

\section{Datumless Measures}

The datumless measures describe various aspects about the relief of a point relative to its surroundings.

\subsection{Dominance}

The \textit{dominance} of point \(\Point{p}\) is the maximum height of \(\Point{p}\) above the horizontal plane of any point on the planetary surface:
\[
\Dominance{p} = \max_{\Point{q} \, \in \, \Manifold{s}} \, \Height{p}{q}
\]

Dominance measures how high a point rises above its large-scale surroundings.\footnote{After dominance was first introduced to the public, it was made known to the author that mountaineers Jerry Brekhus and Andy Martin thought of a concept nearly identical to dominance back in 2006 in an online forum, which they deemed "curvature-adjusted prominence." The only minor difference is in the exact way that the horizontal level is defined.} It is guaranteed to be greater than equal to 0 for any point on the planetary surface (unlike elevation, which may be negative). This is because the height of \(\Point{p}\) above the horizontal plane of itself is 0, and dominance involves taking a maximum with respect to all surface points, including \(\Point{p}\) itself. In addition, due to planetary curvature, \(\Point{p}\) has a negative height with respect to points very far away. Hence, only points within a certain vicinity of \(\Point{p}\) are relevant to the calculation of dominance.

The \textit{base} of point \(\Point{p}\) is the point on the planetary surface that maximizes the height of \(\Point{p}\) above its horizontal plane. The height of \(\Point{p}\) above the horizontal plane of its base is in turn equal to the dominance of \(\Point{p}\). For a point within a mountain range, its base is typically located where the mountain range meets lower plains. The dominance of a mountain's summit thus provides a non-arbitrary base-to-peak measure of the mountain's height.

On Earth, most points have a base close to sea level, therefore measuring a dominance that is usually only slightly lower than elevation. However, for points with an elevated base, usually on a high plain or plateau, dominance can be significantly lower than elevation, providing a more perceptually accurate measure of height. Consider the summit of Pikes Peak in the Front Range of Colorado. The elevation of the summit is \SI{4352}{\meter}, a value that correlates well with the air pressure, climate, and vegetation of the peak. In contrast, its dominance is \SI{2575}{\meter}, describing how much it rises above the neighboring Great Plains. Likewise, a point on the Great Plains may have an elevation above \SI{1000}{\meter}, but its dominance will be close to 0, reflecting the sheer flatness of the surroundings.

\begin{figure}[H]
\centering
\includegraphics[width=\columnwidth]{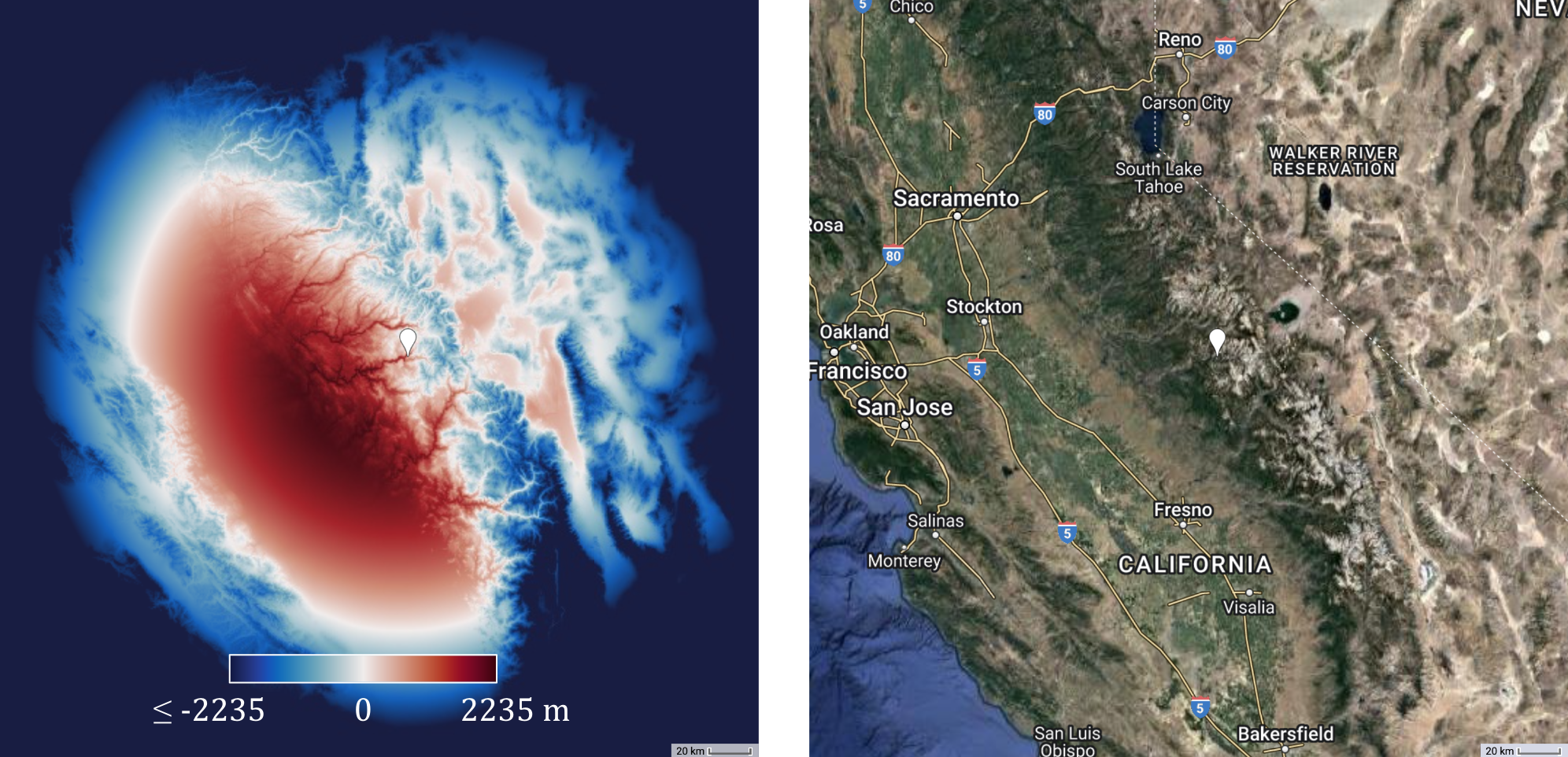}
\caption{
The left map (to be overlaid on the right map) shows the height of the summit of Half Dome (shown as pin) above the horizontal planes of various points in its surroundings. The dominance of the summit equals the maximum value, \SI{2235}{\meter}, measured from its base at the bottom of the Sierra Nevada mountain range where it meets the Central Valley. Values are negative far away due to planetary curvature. Maps are generated in Google Earth Engine \parencite{google-earth-engine}.
}
\end{figure}

A point with a dominance equal to (or less than) 0 is known as a \textit{submissive point}. A submissive point does not rise above the horizontal plane of any point on the planetary surface. Such points are usually found in recessed features such as valleys, canyons, trenches, and craters. An example of a submissive point is Badwater Basin in Death Valley.

\subsection{Jut}

The \textit{jut} of point \(\Point{p}\) is the maximum angle-reduced height of \(\Point{p}\) above any point on the planetary surface:
\[
\Jut{p} = \max_{\Point{q} \, \in \, \Manifold{s}} \, \AngleReducedHeight{p}{q}
\]

Jut measures how sharply or impressively a point rises above its immediate surroundings, yielding a greater value for higher and steeper rises. It is greater than or equal to 0 for any point on the planetary surface. Jut is inspired by the omnidirectional relief and steepness (ORS) measure \parencite{spire-measure}, which was designed to capture the visual impressiveness of a mountain peak or other protruding landform. Jut achieves the same objective in a much simpler way, as it only requires taking a maximum, as opposed to a surface integral in the case of ORS.

The point on the planetary surface that measures the maximum angle-reduced height of \(\Point{p}\) above itself is known as the \textit{immediate base} of \(\Point{p}\). For a point within a mountain range, its immediate base is usually at the bottom of a major mountain face or a neighboring valley (as opposed to its base, which is usually at the bottom of a mountain range). One can find the height of \(\Point{p}\) above the horizontal plane of its immediate base, as well as the angle of elevation of \(\Point{p}\) above its immediate base, thus providing non-arbitrary definitions of base-to-peak height and base-to-peak steepness.

\begin{figure}[H]
\centering
\includegraphics[width=\columnwidth]{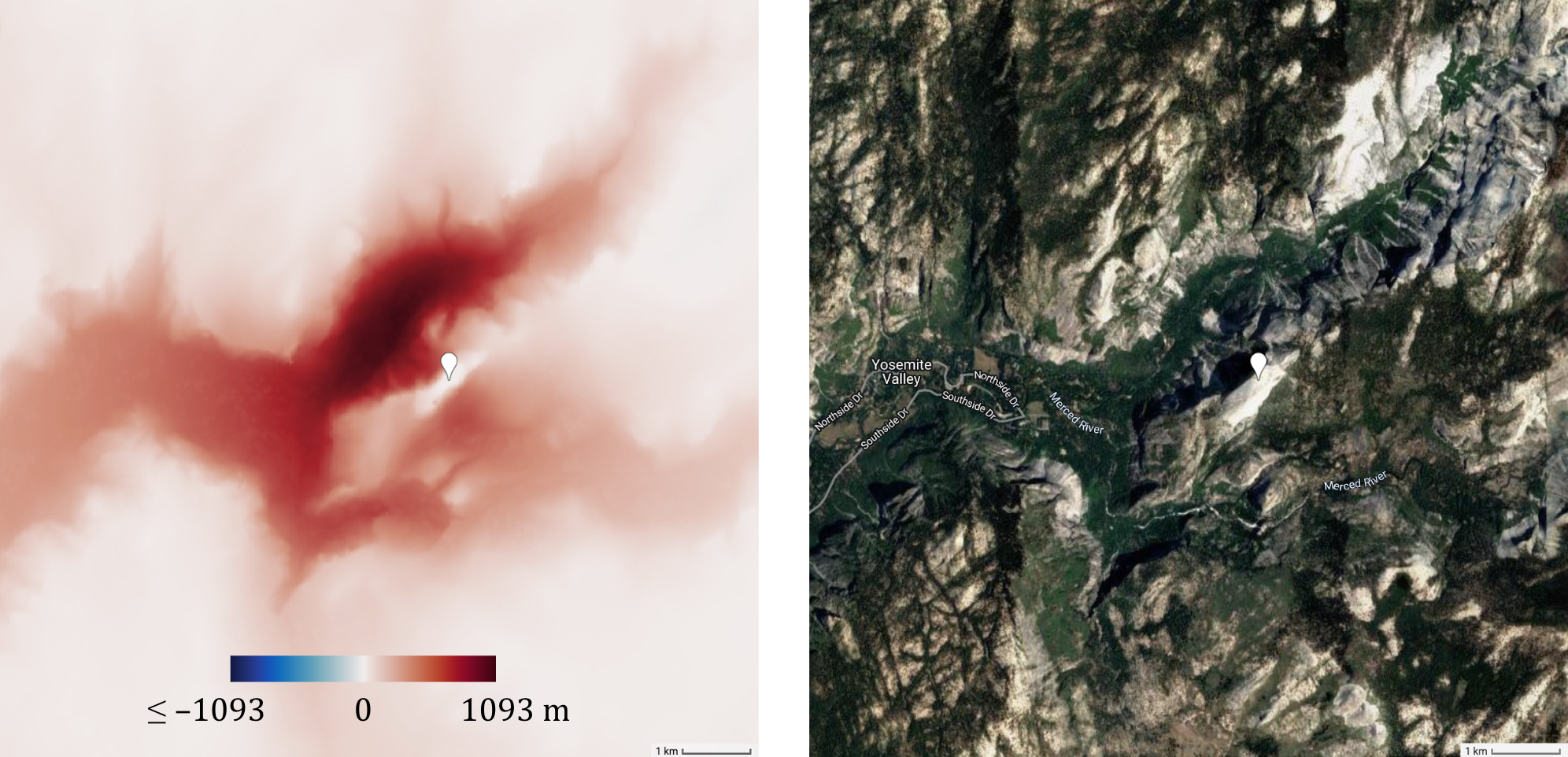}
\caption{
Map showing the angle-reduced height of the summit of Half Dome above points in its immediate surroundings. The jut of the summit equals the maximum value, \SI{1093}{\meter}, measured from its immediate base in Yosemite Valley near Mirror Lake.
}
\end{figure}

\begin{figure}[H]
\centering
\frame{\includegraphics[width=0.48\columnwidth]{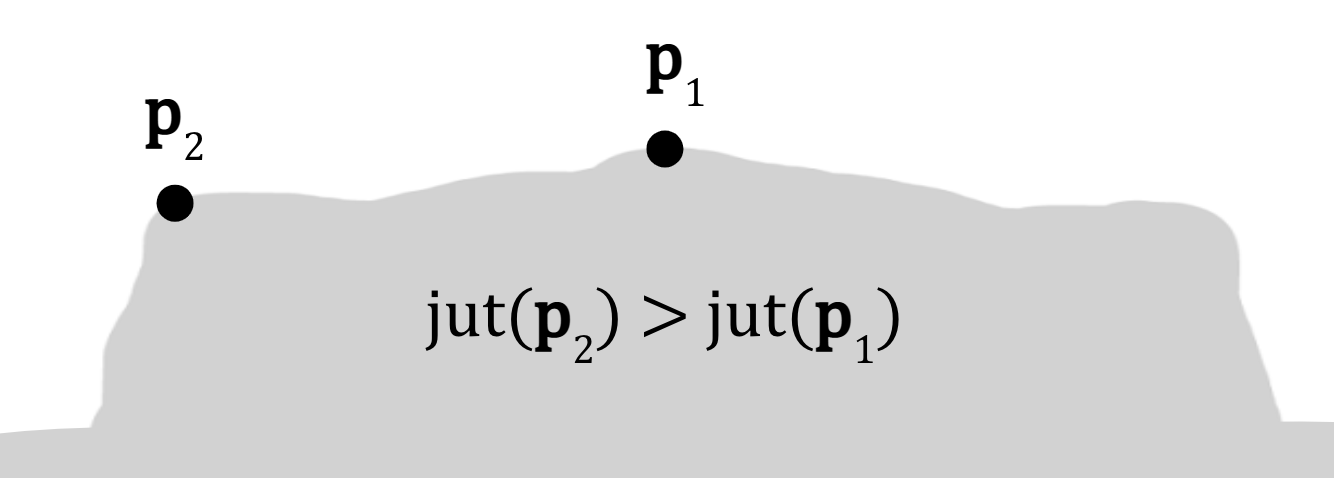}}
\caption{
The point on a mountain with the highest jut is not necessarily its peak. A real-world example is El Capitan in Yosemite, whose maximum jut is measured near the edge of its cliff rather than at higher locations further away from the cliff.
}
\end{figure}

\subsection{Submission}

The \textit{submission} of point \(\Point{p}\) is the maximum height of any point on the planetary surface above the horizontal plane of \(\Point{p}\):
\[
\Submission{p} = \max_{\Point{q} \, \in \, \Manifold{s}} \, \Height{q}{p}
\]

Submission measures how high the surroundings of a point rise above the point itself, yielding a value greater than or equal to 0 for any point on the planetary surface. As with dominance, submission only considers points within a local vicinity, as points very far away from \(\Point{p}\) correspond to negative height values irrelevant to the calculation of submission.

The point on the planetary surface with the maximum height above the horizontal plane of \(\Point{p}\) is known as the \textit{roof} of \(\Point{p}\). For a point within a mountain range, its roof is usually located at the top of a high mountain within the mountain range.\footnote{In the spirit of submission, one can also define an \textit{angular submission} measure that equals the maximum angle of elevation of any point on the planetary surface above \(\Point{p}\). Such a measure would apply strictly to mountain peaks, and can quantify the degree of obstruction of the summit views.}

A point with a submission equal to (or less than) 0 is known as a \textit{dominant point}. A person standing at a dominant point is ``on top of the world,'' as no point rises above their horizontal plane. Dominant points are usually found on protruding features such as mountains, hills, and islands. An example of a dominant point is the summit of Mount Whitney, the highest point of the Sierra Nevada in terms of both dominance and elevation.\footnote{Technically, it is physically possible for a point with a lower elevation or geopotential than \(\Point{p}\) to still be above the horizontal plane of \(\Point{p}\). For mathematical consistency, it might be beneficial to construct a slightly looser definition of a dominant point specifically for mountain peaks. Let a peak \(\Point{p}\)---a local maxima of elevation or geopotential---be considered a \textit{dominant peak} if and only if no point on the planetary surface with a greater geopotential/elevation than \(\Point{p}\) rises above the horizon of \(\Point{p}\). For even more mathematical specificity, one can define a \textit{strictly dominant} point or peak \(\Point{p}\) as one such that ``every point is below the horizontal plane,'' rather than ``no point rises above the horizontal plane.'' A similar problem and solution can be applied to submissive points such as basins.}

\begin{figure}[H]
\centering
\includegraphics[width=\columnwidth]{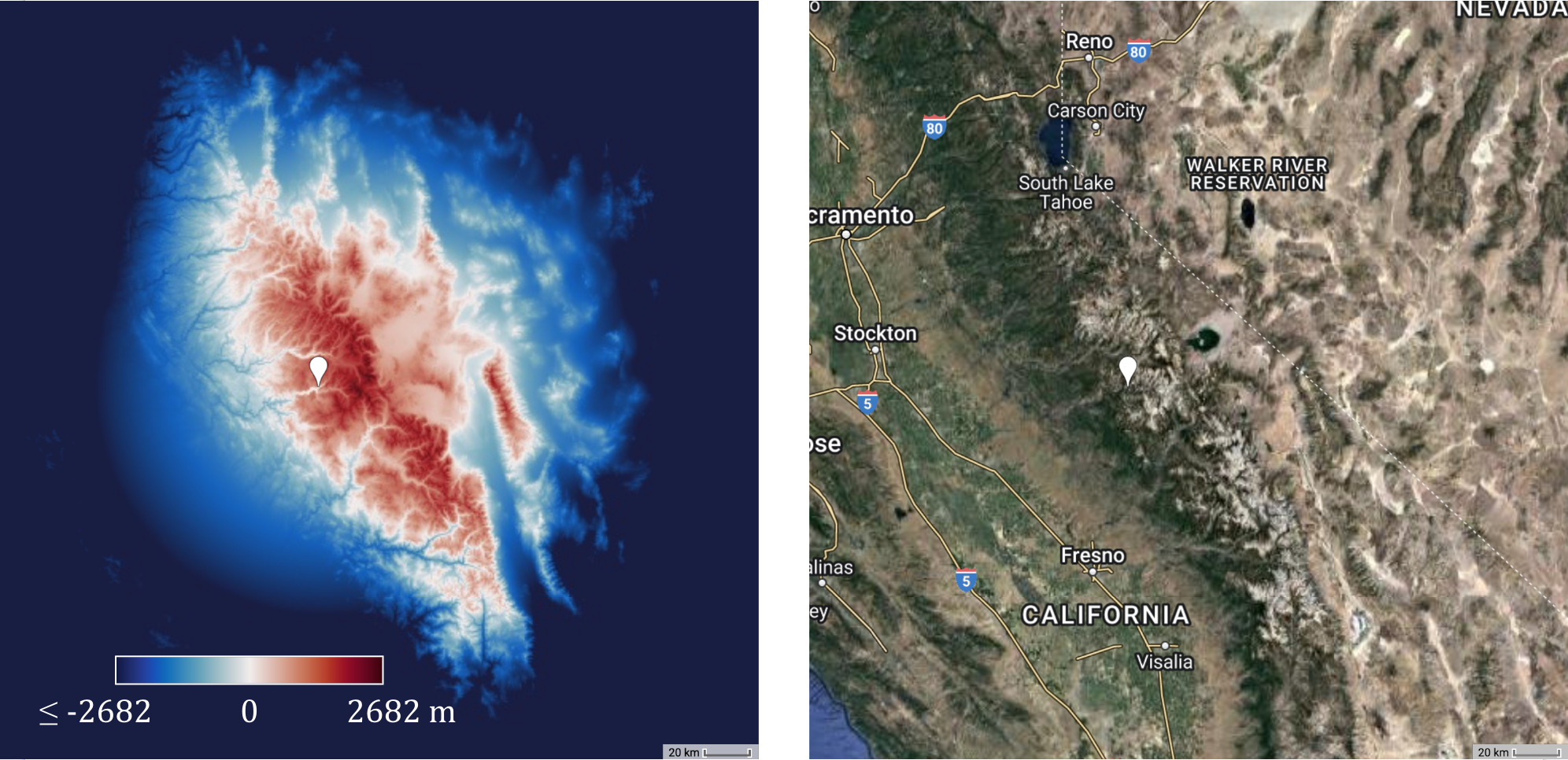}
\caption{
Map showing how high the surroundings of Mirror Lake in Yosemite Valley rise above its horizontal plane. The submission of Mirror Lake equals the maximum value, \SI{2682}{\meter}, measured at its roof at the top of Mount Lyell, the highest mountain in Yosemite National Park. Values are negative far away due to planetary curvature.
}
\end{figure}

\subsection{Rut}

The \textit{rut} of point \(\Point{p}\) is the maximum angle-reduced height of any point on the planetary surface with respect to \(\Point{p}\):
\[
\Rut{p} = \max_{\Point{q} \, \in \, \Manifold{s}} \, \AngleReducedHeight{q}{p}
\]

Rut measures how sharply or impressively the surroundings of a point rise above the point itself, considering both the height and angle of elevation of the surroundings above the point. It is greater than or equal to 0 for any point on the planetary surface.

Whereas jut can be used to quantify the impressiveness of a mountain, rut can be used to quantify how impressively the surrounding mountains rise above a location. That makes it useful making for identifying cities and other non-mountain locations with impressive mountain backdrops.

The point on the planetary surface with the maximum angle-reduced height above \(\Point{p}\) is known as the \textit{immediate roof} of \(\Point{p}\). The immediate roof corresponds to the mountaintop that rises the most sharply or impressively above a location. For a point within a mountain range, its immediate roof is usually at the top of a neighboring mountain (as opposed to its roof, which is usually at the top of a superlative mountain in the mountain range). Knowing the location of the immediate roof, one can find how high it rises above the horizon of \(\Point{p}\), as well as the angle of elevation at which it does so.

\begin{figure}[H]
\centering
\includegraphics[width=\columnwidth]{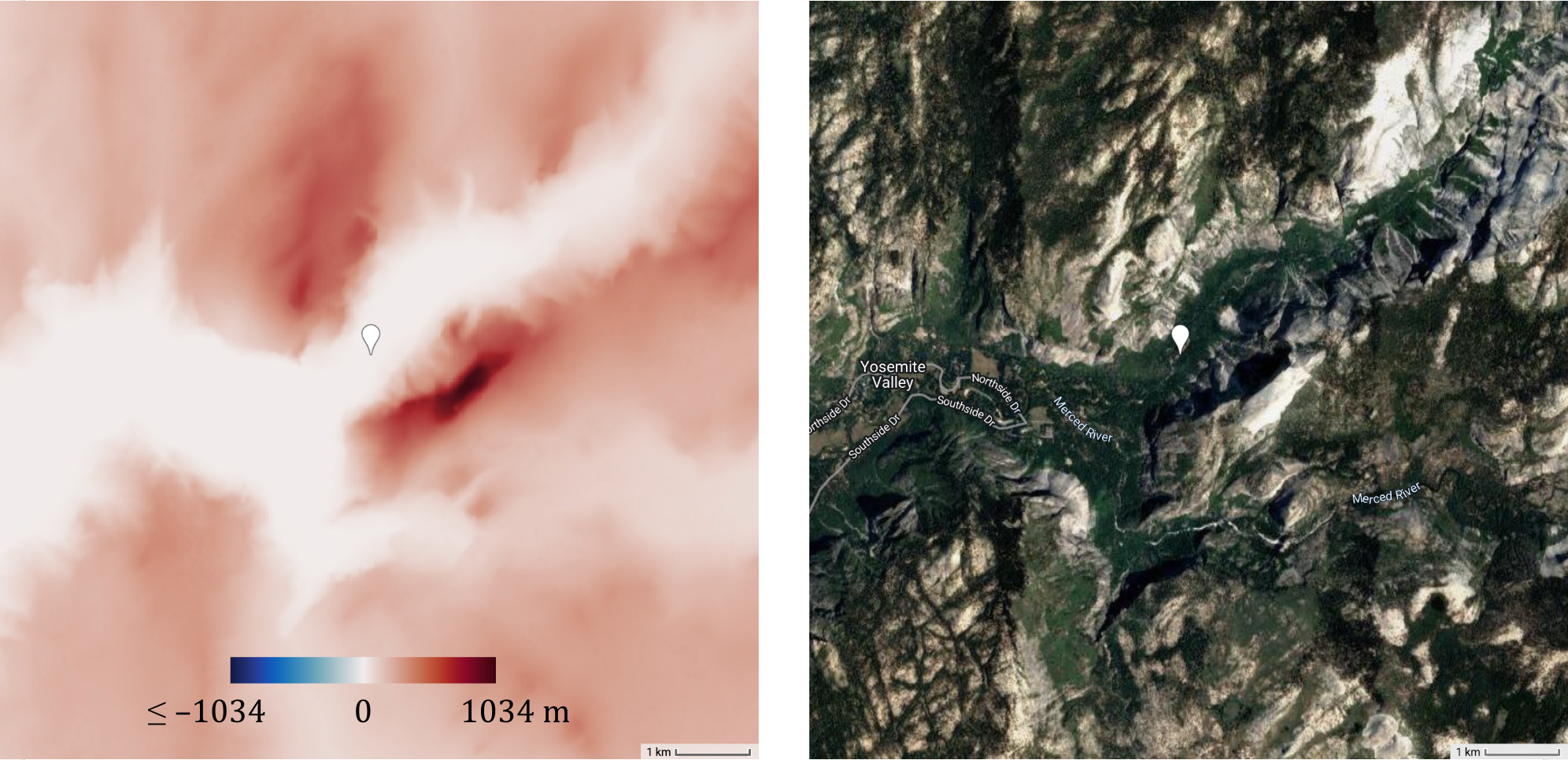}
\caption{
Map showing the angle-reduced height of the surroundings of Mirror Lake above Mirror Lake. The rut of Mirror Lake equals the maximum value, \SI{1034}{\meter}, measured at its immediate roof at the top of Half Dome.
}
\end{figure}

\begin{figure}[H]
\centering
\frame{\includegraphics[width=0.6\columnwidth]{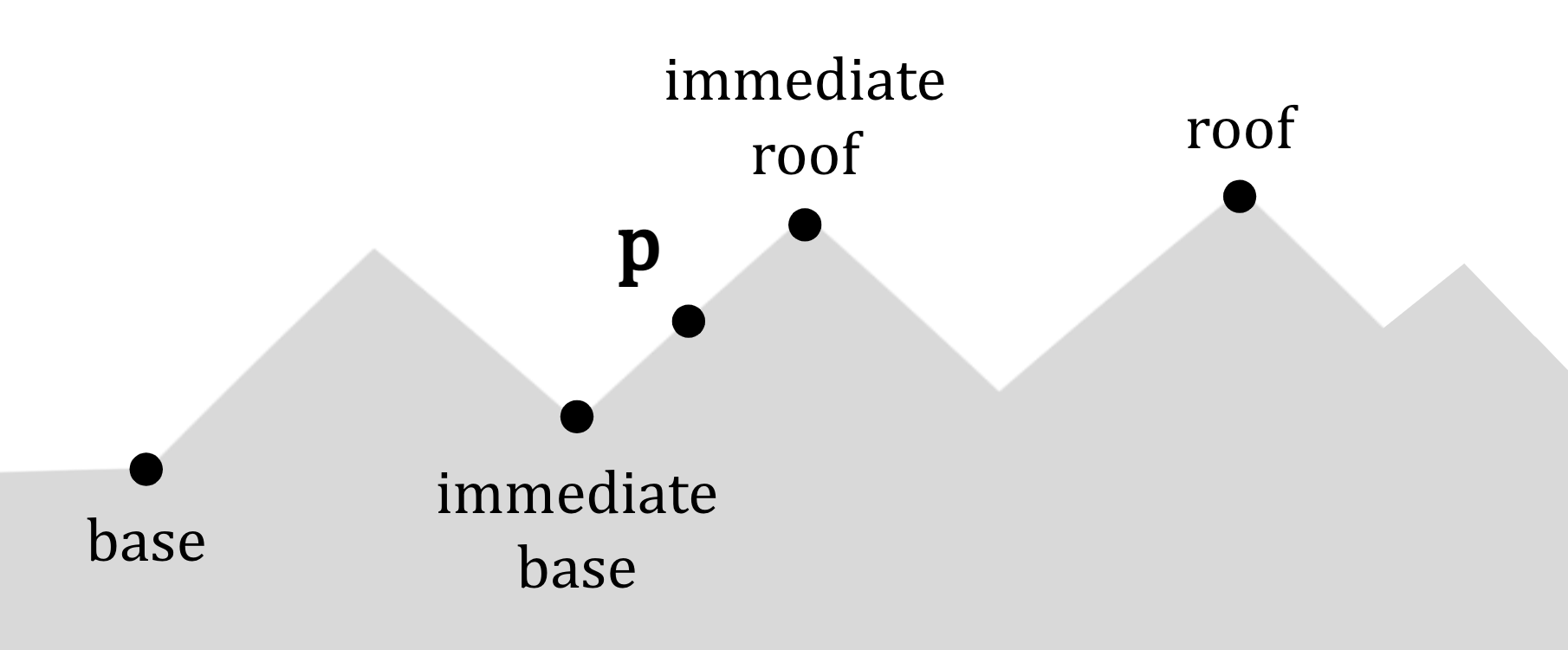}}
\caption{
In this simplified landscape, the base and roof of \(\Point{p}\) are the lowest and highest points of the mountain range that \(\Point{p}\) is in, while its immediate base and immediate roof are the lowest and highest points of the mountain that \(\Point{p}\) is on.
}
\end{figure}

\subsection{Additional Properties of the Datumless Measures}

The datumless measures can be combined to describe even more aspects of relief. By adding up dominance and submission, one gets an indicator of the total amount of relief in the large-scale surroundings of a point. By adding up jut and rut, one gets an indicator of the total degree of vertical fluctuation in the immediate surroundings of a point. One can also use jut-to-dominance ratio to describe how sharply a point gains its immediate relief relative to its large-scale surroundings. On Earth, jut-to-elevation ratio can be used for a similar purpose.\footnote{One can average the datumless measures over a region to quantifying terrain ruggedness. For a less computationally expensive approach, one could take a sample of points within a region.}

Furthermore, one should note that submission is essentially dominance in the opposite direction, and rut is essentially jut in the opposite direction. On a given planet, the base of the point with the highest dominance is the point with the highest submission, and the roof of the point with the highest submission is the point with the highest dominance. Likewise, the immediate base of the point with the highest jut is the point with the highest rut, and the immediate roof of the point with the highest rut is the point with the highest jut.

\section{Computing the Datumless Measures on Earth and Beyond}

This section introduces formulas that can be directly applied in a GIS program to compute the datumless measures. It also showcases the values of the datumless measures at various locations on Earth, Moon, Mars, and Vesta, comparing their values with elevation and prominence.

\subsection{Computing the Datumless Measures}

\subsubsection{Converting a Surface Model to ECEF}

The first step in computing the datumless measures is to convert an existing surface model such as a DEM to an ECEF coordinate system. In this paper, the following models are used for their respective planets:

\begin{table}[H]
\caption{Surface models used for their respective planets.}
\centering
\resizebox{\columnwidth}{!}{%
\begin{tabular}{|l|l|l|l|}
\hline
\textbf{\textbf{Planet}} & \textbf{Model} & \textbf{Height Type} & \textbf{Resolution} \\ \hline
Earth (land and water surface) & ALOS World 3D \parencite{alos-1, alos-2, alos-3, alos-4} & Orthometric      & \SI{30}{\meter} \\
Earth (bathymetry)             & GEBCO\_\,2022 Grid \parencite{gebco}                     & Orthometric      & \SI{450}{\meter}    \\
Earth (geoid)                  & EGM96 \parencite{egm96-1, egm96-2}                       & Geoid            & \SI{450}{\meter}    \\
Moon                           & LRO LOLA \parencite{lola}                                & Ellipsoidal        & \SI{118}{\meter}    \\
Mars (orthometric)                           & MOLA - HRSC Blended \parencite{mola-hrsc}                     & Orthometric            & \SI{200}{\meter}    \\
Mars (areoid)                           & GMM-2B \parencite{gmm-2b}                     & Areoid            & \SI{200}{\meter}    \\
Vesta                          & Dawn FC HAMO \parencite{hamo}                            & Radius           & \SI{93}{\meter}     \\ \hline
\end{tabular}%
}
\end{table}

On Earth and Mars, elevation is presented as orthometric height, or height above the geoid---namely, the EGM96 geoid on Earth and the GMM-2B areoid (essentially a Martian geoid with an arbitrarily defined geopotential) on Mars. Before converting to ECEF, it is a good practice to convert orthometric height to height above a reference ellipsoid---the WGS84 reference ellipsoid on Earth and a sphere with a \SI{3396190}{\meter} radius on Mars.\footnote{The usage of a reference ellipsoid in this case is non-arbitrary, as it is for approximating the direction of gravity---a meaningful physical concept---rather than for providing an arbitrary datum to quantify relief.} The following equation is used for this conversion:
\[h = H + N\]
where \(h\) denotes ellipsoidal height, \(H\) denotes orthometric height, and \(N\) denotes geoid height, i.e., the height of the geoid above the reference ellipsoid. The EGM96 geoid model and GMM-2B areoid model\cite{gmm-2b} are used for \(N\) on Earth and Mars, respectively. This conversion on Mars is done with the help of Ames Stereo Pipeline \parencite{ames-stereo-pipeline}.

The next step is converting ellipsoidal height to \(X\), \(Y\), and \(Z\) coordinates in ECEF using the following equations \parencite{geodetic-to-ecef}:
\begin{align*}
& \latitude = \text{latitude} \\
& \longitude = \text{longitude} \\
& h = \text{ellipsoidal height} \\
& a = \text{length of equatorial radius of reference ellipsoid} \\
& b = \text{length of polar radius of reference ellipsoid} \\
& f = 1 - \frac{b}{a}\;\text{, the flattening of the ellipsoid} \\
& e^2 = 2f - f^2 \\
& N(\latitude) = \dfrac{a}{\sqrt{1 - e^2 \sin^2 \latitude}} \\
& X = (N(\latitude) + h) \cos \latitude \cos \longitude \\
& Y = (N(\latitude) + h) \cos \latitude \sin \longitude \\
& Z = \qty((1 - e^2) \, N(\latitude) + h) \sin \latitude
\end{align*}
The values of \(a\) and \(f\) used for their respective planets are listed in the table below; \(b\) is derivable from \(a\) and \(f\). Heights in the Vesta model, provided as radius from the center of mass, can be treated as height above an ellipsoid with dimensions \(a = 0\) and \(f = 0\).
\begin{table}[H]
\centering
\caption{Values of \(a\) and \(f\) used for their respective planets.}
\begin{tabular}{|l|l|l|}
\hline
\textbf{Planet} & \(a\)                & \(f\)                 \\ \hline
Earth           & \SI{6378137}{\meter} & \(1 \,/\, 298.257223563\) \\
Moon            & \SI{1737400}{\meter} & 0                     \\
Mars            & \SI{3396190}{\meter} & 0                     \\
Vesta           & \SI{0}{\meter}       & 0                     \\ \hline
\end{tabular}
\end{table}

\subsubsection{Approximating the Vertical Unit Vector}

After converting a DEM to ECEF, the next step is to compute the vertical unit vector for points on the planetary surface.

On Earth and Mars, the vertical unit vector can be approximated to point normal to their respective geoid and areoid models, which are created to model an equipotential surface. Using a GIS software, one can compute the angles for the slope \(\slope\) and aspect \(\aspect\) of a geoid or areoid model. (Aspect is an angle from \(\ang{0}\) to \(\ang{360}\) describing the direction that the slope faces, with \(\ang{0}\) denoting a north-facing slope and an increasing angle corresponding to a clockwise rotation on the compass rose.) Given the slope and aspect of the geoid or areoid at a particular latitude-longitude pair, the vertical unit vector at that location is calculated as follows:
\begin{align*}
& \Rotation{z}{\longitude} \times \Rotation{y}{-\latitude} \times \Rotation{x}{-\aspect} \times \Rotation{y}{-\slope} \times\!
\begin{pmatrix}
1 \\ 0 \\ 0
\end{pmatrix} \\
& = \begin{pmatrix}
(\cos\slope\cos\latitude - \sin\slope\cos\aspect\sin\latitude)\cos\longitude - \sin\slope\sin\aspect\sin\longitude \\
(\cos\slope\cos\latitude - \sin\slope\cos\aspect\sin\latitude)\sin\longitude + \sin\slope\sin\aspect\cos\longitude \\
\cos\slope\sin\latitude + \sin\slope\cos\aspect\cos\latitude
\end{pmatrix}
\end{align*}
where the \(R\)\,'s denote 3-dimensional rotation matrices.

On the Moon, where a geoid model not readily available, the vertical unit vector can be approximated to point normal to the reference ellipsoid. Since the reference ellipsoid is a perfect sphere on both planets, the vertical unit vector also points directly away from the planet's center of mass. Given the latitude and longitude of a point, its vertical unit vector normal to the reference ellipsoid is calculated as follows:
\[
\vu{k} = \Rotation{z}{\longitude} \times \Rotation{y}{-\latitude} \times\!
\begin{pmatrix}
1 \\ 0 \\ 0
\end{pmatrix} \\
= \begin{pmatrix}
\cos \latitude \cos \longitude \\
\cos \latitude \sin \longitude \\
\sin \latitude
\end{pmatrix}
\]

On Vesta, a reference ellipsoid of dimensions \(a \!=\! \SI{285}{\kilo\meter}\) and \(b \!=\! \SI{229}{\kilo\meter}\) has been previously used by NASA \cite{vesta-reference-ellipsoid}. However, there lacks a convenient way to convert distance from the center of mass (as provided by the Vesta surface model) to latitude and longitude on the NASA ellipsoid. Therefore, the vertical unit vector on Vesta is approximated with a different approach: Consider the existence of concentric ellipsoids with the same center and the same proportions (i.e., the same \(a \,/\, b\) ratio) as the NASA ellipsoid. The vertical unit vector of a point is approximated as pointing normal to the concentric ellipsoid that it touches. Given the \(X\), \(Y\), and \(Z\) coordinates of a point on Vesta in ECEF, its vertical unit vector \(\vu{k}\) is approximated as follows:
\begin{equation*}
\begin{gathered}
\begin{aligned}
& \Vector{k} =
\begin{pmatrix}
X \,/\, a^2 \\
Y \,/\, a^2 \\
Z \,/\, b^2
\end{pmatrix} \\
& \vu{k} = \frac{\Vector{k}}{\abs{\Vector{k}}}
\end{aligned}
\end{gathered}
\end{equation*}

After computing the ECEF coordinates and vertical unit vectors of points on the planetary surface, standard GIS operations can be used to compute all other quantities related to the datumless measures. To save computational time, the aforementioned computations only need to be applied to points within a certain radius of \(\Point{p}\), as points very far away are irrelevant to the calculation of the datumless measures.

\subsection{Datumless Measures on Earth and Beyond}

Google Earth Engine \parencite{google-earth-engine} is used to compute all values of the datumless measures in this paper, along with generating all maps. Measured points are located with the help of Google Maps \parencite{google-maps}. Summit elevation and prominence values on Earth are either found on Peakbagger.com \parencite{peakbagger} or derived manually from a DEM, with the exception of the dry prominence of Mauna Kea, which is found in \cite{mauna-kea}. Measurements and related conclusions are subject to change and should always be verified.

\subsubsection{Earth}

Earth's topography is unique among the planets in the Solar System. Unlike most planets, its unique plate tectonics play a significant role in forming mountain ranges.

An interesting global phenomenon to note is that mountains closer to polar latitudes tend to have a higher jut-to-elevation ratio, likely due to the effects of glacial erosion in sculpting steeper mountains \parencite{glacier}. Closer to the poles, lower temperatures allow steep, glacier-sculpted flanks to extend to lower terrain, often resulting in immediate bases positioned close to the bottom of a mountain range. Meanwhile, closer to the equator, glaciers are only supported in high altitudes, usually resulting in immediate bases positioned higher in a mountain range.

In the contiguous U.S., the highest dominance values are found in the Sierra Nevada and the Cascade Range, with a handful of summits measuring a dominance above \(\SI{3500}{\meter}\) and Mount Rainier being the only to measure a dominance above \SI{4000}{\meter}. Despite having a similar elevation, the Rocky Mountains have a lower dominance as a result of rising from higher plains, with the greatest values in the American Rockies only slightly exceeding \SI{2500}{\meter}. Places with a jut exceeding \SI{1000}{m} include the Teton Range, Glacier National Park, Yosemite National Park, Mount San Jacinto, the North Cascades, and last but not least, Mount Rainier, which has the highest dominance and jut\footnote{For convenience, the jut of a mountain will refer to its summit measurement, even though non-summit locations may measure a higher jut.} of any major summit\footnote{A \textit{major summit} hereby refers to a peak with at least \SI{300}{\meter} of prominence.} in the contiguous United States. Within the Rocky Mountains, the subranges of Colorado are generally less steep than their more northerly neighbors, with the most impressive\footnote{The term \textit{impressive} is used objectively to denote a high jut.} summits in the state measuring a jut between \SI{500}{\meter} and \SI{850}{\meter}. In comparison to all of these places, the Appalachian Mountains feature significantly less relief as a result of old age, with all locations measuring below \SI{2000}{\meter} of dominance and \SI{500}{\meter} of jut.

In the rest of North America, the Canadian Rockies measure a similar dominance as their American counterparts but a significantly higher jut and jut-to-elevation ratio. Jut values of over \SI{1500}{\meter} are common in the region, culminating at Mount Robson, which has over \SI{1900}{\meter} of jut. The highest and most impressive features in North America are found in Alaska and Northwest Canada, with Denali measuring the highest dominance of over \SI{5500}{\meter} and its North Peak measuring the highest jut of over \SI{2500}{\meter}. Mountains in Mexico tend to have a lower jut-to-elevation ratio, likely due in part to lower glaciation. The highest jut values in the country just barely exceed \SI{1000}{\meter}.

South America is home to the Andes, the highest mountain range outside of Asia in terms of both dominance and elevation. Within the Andes, a handful of major summits measure a dominance of over \(\SI{5500}{\meter}\), with Aconcagua measuring the greatest dominance of over \(\SI{6000}{\meter}\). The greatest jut values in the Andes just exceed \SI{1900}{\meter}. Mountains in the Southern Andes measure the highest jut-to-elevation ratios, likely due to glaciation at lower altitudes as a result of the colder climate. In fact, the jut of major summits in Patagonia are comparable to those of the Central and Northern Andes, which have almost twice the elevation. Meanwhile, mountains in the Central Andes, corresponding to the Atacama Desert and the Altiplano, tend to measure the lowest jut-to-elevation ratios in the entire mountain range, which could in part be due to aridity limiting the formation of glaciers \cite{glacier-2}.

In Africa, mountains tend to measure a lower jut and jut-to-elevation ratio than most other continents, likely due in part to decreased glaciation. The highest and most impressive mountain in the continent is Kilimanjaro, with a dominance of just over \SI{5000}{\meter} and a jut of just over \SI{1300}{\meter}. A few other major summits in Africa measure a dominance of over \SI{4000}{\meter}. A mountain that demonstrates the sculpting effects of glaciation particularly well is Mount Kenya. The extinct stratovolcano rises very gradually from its base, all the way up to a height that supports glaciers, where it then rises sharply to the summit. The immediate base of Mount Kenya is located right where this transition occurs, while its base located where the mountain's lower, more gradual slopes meet flat plains.

In Europe, the two highest mountain ranges are the Alps and Caucasus Mountains. Within the Alps, several major summits exceed \SI{4000}{\meter} of dominance, with Mont Blanc measuring the highest dominance of just over \SI{4400}{\meter}. The Alps have a remarkably high jut-to-elevation ratio, with numerous locations measuring over \SI{1000}{\meter} of jut and a several clusters of peaks measuring over \SI{1500}{\meter}, capping off at the Jungfrau, which measures over \SI{1800}{\meter} of jut. Compared with the Alps, the Caucasus Mountains have slightly higher dominance values, with a few mountains measuring between \SI{4500}{\meter} and \SI{5000}{\meter}, including Mount Elbrus, the mountain with the highest dominance in Europe. However, jut values in the Caucasus are also lower than in the Alps, with the most impressive mountains measuring between \SI{1000}{\meter} and \SI{1600}{\meter} of jut.

Asia is home to the most rugged terrain on Earth. The greatest dominance and jut values on the planet are found in a region bounded roughly by the Tian Shan to the north, the Himalaya to the south, the Hengduan Mountains to the east, and the Pamir-Alay to the west. Within this region, numerous major summits have a dominance of over \SI{6000}{\meter} and a jut of over \SI{2000}{\meter}. The Himalaya measures the highest dominance values, which exceed \SI{7000}{\meter} at a handful of major summits. Mount Everest measures the highest dominance on the planet with a value of \SI{8081}{\meter}, measured from its base at the bottom of the Himalaya where it meets the Indo-Gangetic Plain. Second and third place for major summits with the highest dominance are Kangchenjunga and Lhotse. Meanwhile, K2 barely exceeds \SI{5500}{\meter} of dominance, as the Karakoram mountain range sits on a high plateau. However, both the Himalaya and Karakoram are home to summits exceeding \SI{2500}{\meter} of jut. The highest jut on Earth goes to Annapurna Fang, a subpeak of the Annapurna Massif with a jut of approximately \SI{3400}{\meter}, measured from its immediate base at the bottom of its massive Southwest Face. Other mountains with over \SI{3000}{\meter} of jut include Nanga Parbat, Dhaulagiri, Machapuchare, and Gyala Peri. Asia is also home to the Tibetan Plateau, which is so large that the dominance of mountains in its central regions are measured directly from the high elevations of the plateau, rather than from low plains surrounding it.

In Oceania, the mountain with the greatest dominance is Puncak Jaya, with a value of just over \SI{4800}{\meter}. Australia is considerably flat, with Mount Kosciuszko, its highest-dominance feature, measuring a dominance of just below \SI{2000}{\meter}, and all locations measuring below \SI{600}{\meter} of jut, with the exception of Mount Gower on Lord Howe Island, where jut values are only slightly higher. In contrast, neighboring New Zealand features significantly greater relief. The highest dominance is New Zealand is measured atop Aoraki / Mount Cook, with a value of just over \SI{3600}{\meter}. The glacier-carved fjords of Milford Sound have some of the highest jut-to-elevation ratios in the world, with Mitre Peak measuring a jut of over \SI{1300}{\meter}, similar to that of the Matterhorn which has over twice the elevation.

In Antarctica, the Sentinel Range is the highest mountain range, with Vinson Massif measuring the highest dominance of just over \SI{4500}{\meter} and a few locations measuring a jut of over \SI{1500}{\meter}. Moving inland, the ice sheet thickens so gradually that points in its central portions measure a dominance and jut close to 0, despite having thousands of meters of elevation. A similar phenomenon occurs in the center of the ice sheet of Greenland.

When measuring seabed landforms, the dry datumless measures directly describe their relief relative to the ocean floor, unlike elevation. The greatest dry dominance on Earth---a whopping \SI{10200}{\meter}---is measured atop an unnamed seamount approximately \SI{60}{\kilo\meter} SSW of the southernmost point of Guam. Likewise, the Mariana Trench measures the greatest dry submission on Earth. The shield volcano Mauna Kea in Hawaii also has a significant dry dominance of just over \SI{9300}{\meter}, measured from its base on the ocean floor.

\begin{table}[H]
\caption{Datumless measures (unit: meters) at various summits on Earth.}
\centering
\begin{tabular}{|l|l|l|l|l|l|l|l|}
\hline
\textbf{Mountain}                    & \textbf{Mountain Range}         & \textbf{dom} & \textbf{sub} & \textbf{jut} & \textbf{rut} & \textbf{Elev} & \textbf{Prom} \\ \hline
Mount Washington                       & Appalachians                    & 1751         & 0            & 381          & 0            & 1917          & 1874          \\
Pikes Peak                           & Rocky Mountains                 & 2575         & 0            & 517          & 0            & 4301          & 1680          \\
Grand Teton                          & Rocky Mountains                 & 2421         & 0            & 1125         & 0            & 4197          & 1990          \\
Half Dome                            & Sierra Nevada                   & 2235         & 1252         & 1093         & 68           & 2694          & 414           \\
Mount Whitney                          & Sierra Nevada                   & 3955         & 0            & 757          & 0            & 4419          & 3072          \\
Mount Rainier                          & Cascade Range                   & 4193         & 0            & 1266         & 0            & 4392          & 4037          \\
Mount Robson                           & Rocky Mountains                 & 3157         & 0            & 1907         & 0            & 3959          & 2819          \\
Denali                               & Alaska Range                    & 5765         & 0            & 2101         & 0            & 6190          & 6140          \\
Aconcagua                            & Andes                 & 6014         & 0            & 1832         & 0            & 6962          & 6962          \\
Mount Fitz Roy                         & Andes                 & 3200         & 0            & 1776         & 0            & 3405          & 1951          \\
Kilimanjaro                          & East African Rift               & 5067         & 0            & 1367         & 0            & 5895          & 5885          \\
Table Mountain                       & Cape Fold Belt                  & 1085         & 473          & 465          & 3.0          & 1085          & 1055          \\
Matterhorn                           & Alps                            & 3952         & 233          & 1364         & 3.8          & 4476          & 1038          \\
Mont Blanc                           & Alps                            & 4364         & 0            & 1730         & 0            & 4810          & 4697          \\
Mount Elbrus                           & Caucasus Mountains              & 4925         & 0            & 1106         & 0            & 5642          & 4741          \\
Kirkjufell                           & Sn\ae fellsnes Peninsula                            & 472          & 527          & 259          & 52           & 469           & 449           \\
Mount Everest                          & Himalaya                        & 8081         & 0            & 2109         & 0            & 8849          & 8849          \\
Nanga Parbat                         & Himalaya                        & 7043         & 0            & 3166         & 0            & 8125          & 4608          \\
K2                                   & Karakoram                 & 5831         & 0            & 2542         & 0            & 8614          & 4020          \\
Mount Fuji                             & NE Japan Arc                             & 3731         & 0            & 1062         & 0            & 3776          & 3776          \\
Puncak Jaya                          & Sudirman Range                  & 4842         & 0            & 1303         & 0            & 4884          & 4884          \\
Mount Kosciuszko                       & Snowy Mountains                 & 1911         & 0            & 369          & 0            & 2228          & 2228          \\
Vinson Massif                        & Sentinel Range                  & 4584         & 0            & 1161         & 0            & 4892          & 4892          \\
Unnamed (Dry)        & Mariana Arc       & 10284                           & 124          & 1410         & 1.0          & -27          & 148           \\
Mauna Kea (Dry)                      & Hawaiian Volcanoes & 9333       & 0            & 1203         & 0            & 4205          & 9330          \\ \hline
\end{tabular}
\end{table}

\begin{table}[H]
\caption{Datumless measures (unit: meters) at various non-summit locations on Earth.}
\centering
\begin{tabular}{|l|l|l|l|l|l|}
\hline
\textbf{Location} & \textbf{dom} & \textbf{sub} & \textbf{jut} & \textbf{rut} & \textbf{Elev} \\ \hline
San Francisco     & 16           & 991          & 0.2          & 25           & 17            \\
Denver            & 60           & 2476         & 0.3          & 105          & 1627          \\
Mount Sunflower    & 142           & 46          & 1.3          & 0.1          & 1231           \\
Mirror Lake    & 808           & 2682          & 19          & 1034          & 1263           \\
Whitney Portal    & 2123           & 1862          & 134          & 680          & 2552           \\
Crater Lake       & 1281         & 862          & 32           & 89           & 1882          \\
Mather Point      & 1462         & 1060         & 710          & 13           & 2170          \\
Addis Ababa       & 527          & 1075         & 6.0          & 64           & 2293          \\
Kathmandu & 935         & 6147         & 26          & 492         & 1307          \\
Everest Base Camp & 4558         & 3477         & 309          & 1510         & 5364          \\
Lhasa             & 31           & 2728         & 0.6          & 287          & 3656          \\
Challenger Deep (Dry)   & 0            & 8892         & 0            & 1274         & -10923        \\ \hline
\end{tabular}
\end{table}

\subsubsection{Moon, Mars, and Vesta}

The datumless measures are particularly handy on planets without a sea level, as they directly describe relief relative to local terrain (as opposed to elevation, whose values need to be compared with other elevation values for this purpose). Generally, smaller planets tend to feature greater relief as quantified by the datumless measures. This is likely due to the lower gravitational pull allowing higher mountains to form at isostatic equilibrium \parencite{isostasy}.

The Moon is significantly more rugged than the Earth when it comes to averages. The far side of the moon facing away from Earth is more rugged than the near side. The point with the highest elevation, known as the Selenian Summit, measures a dominance of slightly over \SI{8300}{\meter} and a jut of slightly over \SI{1100}{\meter}. However, some places with a lower elevation feature even greater relief, with dominance exceeding \SI{9000}{\meter} and jut exceeding \SI{2000}{\meter} in a few locations on the far side. The greatest dominance and jut values take place at an unnamed summit approximately \SI{200}{\kilo\meter} northwest of the Lippmann Crater, with dominance exceeding \SI{10000}{\meter} and jut exceeding \SI{2600}{\meter}. The near side of the Moon is less rugged than the far side, containing several maria---large, flat impact basins whose central locations typically measure a rut of below \SI{10}{\meter}. The famous Montes Apenninus mountain range near the Apollo 15 landing site measures dominance values above \SI{5000}{\meter} and jut values above \SI{2000}{\meter} in places, with the highest dominance measured atop Mons Huygens.

Mars is a land of topographic extremes, with superlative datumless measurements exceeding those of the Moon. The giant shield volcano Olympus Mons measures the highest dominance on the planet, with its summit measuring a value of just over \SI{14100}{\meter}. Despite this, Olympus Mons rises very gradually; in fact, the summit is below the horizontal plane of some locations at the bottom of the volcano's lower cliffs. Mars is also home to the giant Valles Marineris canyon, with several times the relief of the Grand Canyon, measuring over \SI{10000}{\meter} of submission and over \SI{3500}{\meter} of rut at certain places. Yet another notable feature on Mars is the Hellas Planitia impact basin. Locations near the rim of the basin can measure a submission of over \SI{5500}{\meter} and a rut of over \SI{1000}{\meter}, but points in its central portions typically measure a low rut below \SI{10}{\meter}, indicative of flat surroundings. The northern hemisphere of Mars contains large swaths of plains that also consistently measure rut values below \SI{10}{\meter}.

Of all the planetary bodies mentioned in this paper, Vesta is by far the most rugged. The most salient feature on Vesta is Rheasilvia, an impact crater with a central peak near the south pole. The central peak measures a dominance of over \SI{16000}{\meter}, even greater than that of Olympus Mons. The rim measures even higher dominance values exceeding \SI{18000}{\meter}, and jut values exceeding \SI{6000}{\meter}. The equatorial and northern regions of Vesta are comparatively less rugged, but still very much so. The Divalia Fossa canyon near the equator has less relief than Valles Marineris on Mars, but still significantly more relief than the Grand Canyon. Overall, Vesta is extremely rugged, with even the flattest locations measuring thousands of meters of submission and hundreds of meters of rut.

\begin{table}[H]
\caption{Datumless measures (units: meters) on Moon, Mars, and Vesta.}
\centering
\resizebox{\columnwidth}{!}{%
\begin{tabular}{|l|l|l|l|l|l|}
\hline
\textbf{Planet}        & \textbf{Location}                             & \textbf{dom} & \textbf{sub} & \textbf{jut} & \textbf{rut} \\ \hline
\multirow{7}{*}{Moon}  & Apollo 11 Landing Site                        & 372.8        & 237.3        & 6.8          & 3.1 \\
                       & Apollo 15 Landing Site                        & 773          & 3957         & 23           & 780 \\
                       & Mons Hadley                                   & 4726         & 0            & 1953         & 0 \\
                       & Mons Huygens                                  & 5151         & 0            & 1570         & 0 \\
                       & Selenian Summit                               & 8352         & 0            & 1147         & 0 \\
                       & Unnamed Summit, \(\sim\!\SI{200}{\kilo\meter}\) NW of Lippmann & 10806        & 0            & 2600         & 0 \\
                       & Mare Imbrium, Central                         & 79           & 7            & 0.7          & 0.01 \\ \hline
\multirow{7}{*}{Mars}  & Olympus Mons, Summit                          & 14175        & 0            & 947          & 0 \\
                       & Olympus Mons, NW Clifftop                     & 13361        & 345          & 2836         & 0.5 \\
                       & Valles Marineris, Coprates Chasma, Floor      & 0            & 10379        & 0            & 1922 \\
                       & Valles Marineris, Candor Chasma, Rim          & 8048         & 2310         & 3562         & 67 \\
                       & Hellas Planitia, Central                      & 75           & 414          & 1.1          & 4.5 \\
                       & Curiosity Landing Site                        & 53           & 4808         & 0.4          & 761 \\
                       & Mount Sharp                                     & 4904         & 0            & 1507         & 0 \\ \hline
\multirow{5}{*}{Vesta} & Rheasilvia, Central Peak                      & 16321        & 0            & 3060         & 0 \\
                       & Rheasilvia, Floor                             & 0            & 16916        & 0            & 4108\\
                       & Rheasilvia, Rim                               & 18920        & 287          & 6492         & 11 \\
                       & Divalia Fossa, Floor                          & 0            & 4332         & 0            & 1235 \\
                       & Feralia Planitia, Floor                       & 611          & 8647         & 114          & 2133 \\ \hline
\end{tabular}%
}
\end{table}

\newpage

\section{Acknowledgements}

Thank you to everyone who helped make this paper a reality, and who has been supportive of the ideas in this paper.

Thank you to Tony Wang for revising the mathematical notation to make it clear and concise. You are truly a genius and a wonderful mentor, and it means a lot for this paper to be read by you.

Thank you to Mr.\@ Behrooz Shahrvini for teaching me so many of the mathematical concepts used in this paper. It brings me joy and fulfillment to apply what I learned in your class to ``spicy problems,'' as you would often say.

Thank you to the Yale Undergraduate Research Association for providing the opportunity to present several key ideas in this paper at the Yale Undergraduate Research Symposium.

Thank you to Dr.\@ Mark Brandon for teaching me what a geoid is. I once thought I could write this entire paper without knowing such fundamentals---you gladly proved me wrong.

Thank you to Mrs.\@ Pamela Meuser for kindling my interest in STEM subjects when I was in elementary school. You have unknowingly set the trajectory for many of my current goals and undertakings.

Thank you to John Metcalfe for writing an article in The Mercury News about California cities with impressive mountain backdrops, as quantified by the rut measure.

Thank you to Greg Slayden and Andrew Kirmse for your efforts to incorporate jut measurements onto Peakbagger.com.

Thank you to the people of Reddit who have been supportive of my work and helped me improve the way I present my ideas.

Thank you to all my friends for flying together with me during this journey. I must apologize for my excessive rants about mountains during this weird phase of my life.

Last but not least, thank you to my parents, Helen Kang and Vince Xu, for supporting me through all the ups and downs of the complex topography of life.

\newpage

\section{References}

\printbibliography[heading=none]

\end{document}